\documentclass[twocolumn,aps,prc,floatfix]{revtex4}
\usepackage{graphicx}
\usepackage{dcolumn}
\usepackage{bm}
\usepackage{mhchem}

\begin{document}

\title{Effects of the initialization of nucleon momentum in heavy-ion collisions at medium energies}
\author{Zu-Xing Yang$^{1,2}$}
\author{Xiao-Hua Fan$^{1,2,3}$}
\author{Gao-Chan Yong$^{1,2}$}
\email[]{yonggaochan@impcas.ac.cn}
\author{Wei Zuo$^{1,2}$}
\affiliation{$^1$Institute of Modern Physics, Chinese Academy of Sciences, Lanzhou 730000, China\\
$^2$School of Nuclear Science and Technology, University of Chinese Academy of Sciences, Beijing 100049, China\\
$^3$School of Physical Science and Technology, Lanzhou University, Lanzhou 730000, China}

\begin{abstract}

Based on the Isospin-dependent transport model Boltzmann-Uehling-Uhlenbeck (IBUU),
effects of the difference of the high momentum tails (HMTs) of nucleon momentum distribution in colliding nuclei on some isospin-sensitive observables are studied in the $^{197}\rm {Au}+^{197}\rm {Au}$ reactions at incident beam energy of 400 MeV/nucleon. It is found that the nucleon transverse and elliptic flows, the free neutron to proton ratio at low momenta are all less sensitive to the specific form of the HMT, while the free neutron to proton ratio at high momenta and the yields of $\pi^{-}$ and $\pi^{+}$ as well as the $\pi^{-}/\pi^{+}$ ratio around the Coulomb peak are sensitive to the specific form of the HMT. Combining the present studies with the experimental measurements at rare-isotope reaction facilities worldwide, one may get more insights into the nuclear short-range correlations in heavy nuclei or nuclear matter.

%Keywords:

\end{abstract}

\maketitle

\section{introduction}

Recently, the studies of the nuclear short-range correlations (SRC) have been shown more interests in nuclear community \cite{RMP2017,ppnp2018,Sargsian17,Claudio15,cailichenprc2016,caili2prc2016,ding2016,cuy2016,yongl2016,xuprc2017,
yongl2017,yongprc2016,yongprc2017,yongliprc2017,yongl2018,
sci14,henprc15,Rios09,Rios14,yin13,sci08}.
The basic points of the nuclear short-range correlations are clear, i.e., in the nuclei from  light to heavy, about $20\%$ nucleons form pairs with larger relative momentum and smaller center-of-mass momentum. And there is a high-momentum tail (HMT) of nucleon momentum distribution in nuclei. The nuclear short-range correlations are about 18 times stronger in neutron-proton pairs than in neutron-neutron or proton-proton pairs.

For the few-nucleon systems or light nuclei, the nuclear wave functions and consequently, the nucleon momentum distributions can be obtained \emph{ab initio} with realistic Hamiltonians \cite{Claudio15}. While for the case of heavy nuclei, usually the approximate parameterized phenomenological wave functions are adopted. Therefore, the specific form of the high-momentum tail of nucleon momentum distribution in heavy nuclei or nuclear matter is controversial in the literature \cite{cailichenprc2016,yongliprc2017,Rios14,yin13,henprc15}. The specific fractions of high-momentum nucleons in heavy nuclei are different in experimental and theoretical studies \cite{cailichenprc2016,yongliprc2017,Rios14,yin13,henprc15}.
And also the effective cutoff value of the maximum momentum of nucleon in heavy nuclei or nuclear matter varies in the literature \cite{yongl2017,cailichenprc2016,henprc15}.
Moreover, the transition momentum of minority in asymmetric matter is rarely studied \cite{yongl2018}. These different aspects of the initialization of nucleon momentum in heavy nuclei or nuclear matter may affect the isospin-sensitive observables in heavy-ion collisions at medium energies. And the isospin-sensitive observables in heavy-ion collisions are very useful to constrain the equation of state of asymmetric nuclear matter which plays important roles in nuclear physics and astrophysics \cite{topic}.

In the present study, unlike the general study of the effects of the HMT on free nucleon emission in heavy-ion collisions \cite{yongl2016,xuprc2017}, we try to see if the difference of  the initialization of nucleon momentum in heavy nuclei or nuclear matter really has effects on some isospin-sensitive observables in heavy-ion collisions at intermediate energies. We carry out such studies in the central and semi-central $^{197}\rm {Au}+^{197}\rm {Au}$ reactions at incident beam energy of 400 MeV/nucleon. One finds that the nucleon flow and the free neutron to proton ratio at low momenta are both less sensitive to the specific form of the HMT, while the free neutron to proton ratio at high momenta, the yields of $\pi^{-}$ and $\pi^{+}$ as well as the $\pi^{-}/\pi^{+}$ ratio around the Coulomb peak are all sensitive to the specific form of the HMT. Besides the caution of constraining the nuclear symmetry energy by isospin-sensitive but HMT-insensitive observables, these studies may help one to constrain the specific form of the nucleon momentum distribution above the Fermi momentum and to get more insights into the nuclear short-range interactions in heavy nuclei or nuclear matter.

\section{the initialization of nucleon momentum and the transport model}

In the used IBUU model, nucleon spatial
distribution in initial colliding nuclei is given by \cite{bertsch,yongl2017}
\begin{equation} \label{xyz}
r = R(x_{1})^{1/3}; cos\theta = 1-2x_{2}; \phi = 2\pi x_{3}.
\end{equation}
\begin{equation}\label{1}
    x = rsin\theta cos\phi;
    y = rsin\theta sin\phi;
    z = rcos\theta.
\end{equation}
Where $R$ is the radius of nuclei, $x_{1}, x_{2}, x_{3}$ are
three independent random numbers.
Since the specific form of the high-momentum tail of nucleon momentum distribution in heavy nuclei or nuclear matter is still controversial, in this study the proton and neutron momentum distributions with high-momentum tail are given by the extended Brueckner-Hartree-Fock (BHF) approach by adopting the AV 18 two-body interaction plus a microscopic Three-Body-Force (TBF) with about 15\% nucleons in the HMT \cite{yin13} (labelled by ``IMD1'') or a simple parameterized n-p dominance form with about 20\% nucleons in the HMT \cite{yongprc2017} (labelled by ``IMD2''), as well as the more complicated parameterized form with about 28\% nucleons in the HMT \cite{cailichenprc2016} (labelled by ``IMD3'').
\begin{figure}[th]
\centering
\includegraphics[width=0.5\textwidth]{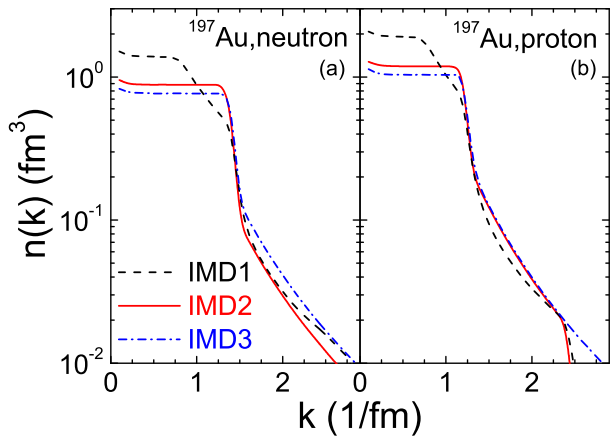}
\caption{ (Color online) Momentum distribution n(k) of neutron (a) and proton (b) in nucleus $^{197}_{79}Au$ given by the BHF with Av18+TBF \cite{yin13} (labelled by ``IMD1''), or a simple parameterized form \cite{yongprc2017} (labelled by ``IMD2''), and the more complicated parameterized form \cite{cailichenprc2016} (labelled by ``IMD3'').
The normalization condition  $\int_{0}^{k_{max}}n(k)k^{2}dk$ = 1.} \label{npdis}
\end{figure}
Fig.~\ref{npdis} shows our used three nucleon momentum distributions in the $^{197}_{79}$Au nucleus. From Fig.~\ref{npdis}, it is seen that more nucleons are in the Fermi sea with the IMD1 while more nucleons are located in the HMT with the IMD3. Around the Fermi momentum, there are more nucleons with the IMD2 and the IMD3. The cutoff values are about 2 (IMD1, IMD2) and 2.38 (IMD3) times the Fermi momentum, respectively.

In this model, the SRC modified isospin- and momentum-dependent mean-field
single nucleon potential is used \cite{yongprc2016,yongprc2017},
which reads
\begin{eqnarray}
U(\rho,\delta,\vec{p},\tau)&=&A_u(x)\frac{\rho_{\tau'}}{\rho_0}+A_l(x)\frac{\rho_{\tau}}{\rho_0}\nonumber\\
& &+B(\frac{\rho}{\rho_0})^{\sigma}(1-x\delta^2)-8x\tau\frac{B}{\sigma+1}\frac{\rho^{\sigma-1}}{\rho_0^\sigma}\delta\rho_{\tau'}\nonumber\\
& &+\frac{2C_{\tau,\tau}}{\rho_0}\int
d^3\,\vec{p^{'}}\frac{f_\tau(\vec{r},\vec{p^{'}})}{1+(\vec{p}-\vec{p^{'}})^2/\Lambda^2}\nonumber\\
& &+\frac{2C_{\tau,\tau'}}{\rho_0}\int
d^3\,\vec{p^{'}}\frac{f_{\tau'}(\vec{r},\vec{p^{'}})}{1+(\vec{p}-\vec{p^{'}})^2/\Lambda^2},
\label{buupotential}
\end{eqnarray}
where $\tau, \tau'=1/2(-1/2)$ for neutrons (protons),
$\delta=(\rho_n-\rho_p)/(\rho_n+\rho_p)$ is the isospin asymmetry,
and $\rho_n$, $\rho_p$ denote neutron and proton densities,
respectively. The SRC modified parameter values $A_u(x)$, $A_l(x)$, B,
$C_{\tau,\tau}$, $C_{\tau,\tau'}$, $\sigma$, and $\Lambda$ can be found
in Ref. \cite{yongprc2016}. In this study, since
the symmetry energy may be mildly soft \cite{yongprc2016}, we let $x$= 1.
The in-medium isospin-dependent baryon-baryon ($BB$) scattering cross section (including elastic and inelastic) $\sigma_{BB}^{medium}$ is reduced compared with the free-space value
$\sigma _{BB}^{free}$ by a factor of
\begin{eqnarray}
R^{BB}_{medium}(\rho,\delta,\vec{p})&\equiv& \sigma
_{BB_{elastic, inelastic}}^{medium}/\sigma
_{BB_{elastic, inelastic}}^{free}\nonumber\\
&=&(\mu _{BB}^{\ast }/\mu _{BB})^{2},
\end{eqnarray}
where $\mu _{BB}$ and $\mu _{BB}^{\ast }$ are the reduced masses
of the colliding baryon-pair in free space and medium, respectively. More details on the particle-particle collisions can be found in Ref. \cite{yongprc2017}.

\section{results and discussions}

\begin{figure}[th]
\centering
\includegraphics[width=0.5\textwidth]{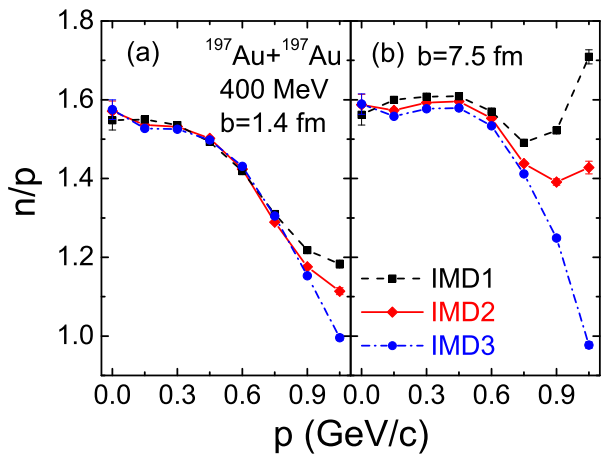}
\caption{(Color online) Free neutron to proton ratio as a function of momentum with different initializations of nucleon momentum (IMD1, IMD2, IMD3) in the $^{197}\rm {Au}+^{197}\rm {Au}$ reactions at incident beam energy of 400 MeV/nucleon with impact parameters b= 1.4 fm (a) and 7.5 fm (b), respectively.} \label{rnp}
\end{figure}
Since the SRC or the HMT is the behavior of nucleons in nuclei or nuclear matter, nucleon observable should be the best probe of the HMT. Fig.~\ref{rnp} shows the free neutron to proton ratio n/p as a function of momentum with different HMTs in the $^{197}\rm {Au}+^{197}\rm {Au}$ reactions at incident beam energy of 400 MeV/nucleon. From Fig.~\ref{rnp} (a), it is seen that the free neutron to proton ratio n/p is less sensitive to the HMT at low momenta. However, the effects of the difference of HMTs become larger at high momenta. Since more neutrons and protons are correlated with the case IMD3 than with the case IMD1, it is not a surprise one sees the value of free n/p with IMD3 is lower than that with IMD1. From Fig.~\ref{rnp}, it is also seen that, for the central collision (b= 1.4 fm), the effects of the HMT can reach about 20\% at very high momenta. While the effects are more clear in the semi-central (b= 7.5 fm) $^{197}\rm {Au}+^{197}\rm {Au}$ reaction. The effects of the difference of the HMTs on the free n/p at very high momenta reach about 50\%, which are of course measurable at rare-isotope reaction facilities worldwide. For the semi-central collisions, since more (less) energetic nucleons are from the HMT (nucleon-nucleon collisions), the larger effects of the HMT on the free n/p ratio are expected.

\begin{figure}[th]
\centering
\includegraphics[width=0.5\textwidth]{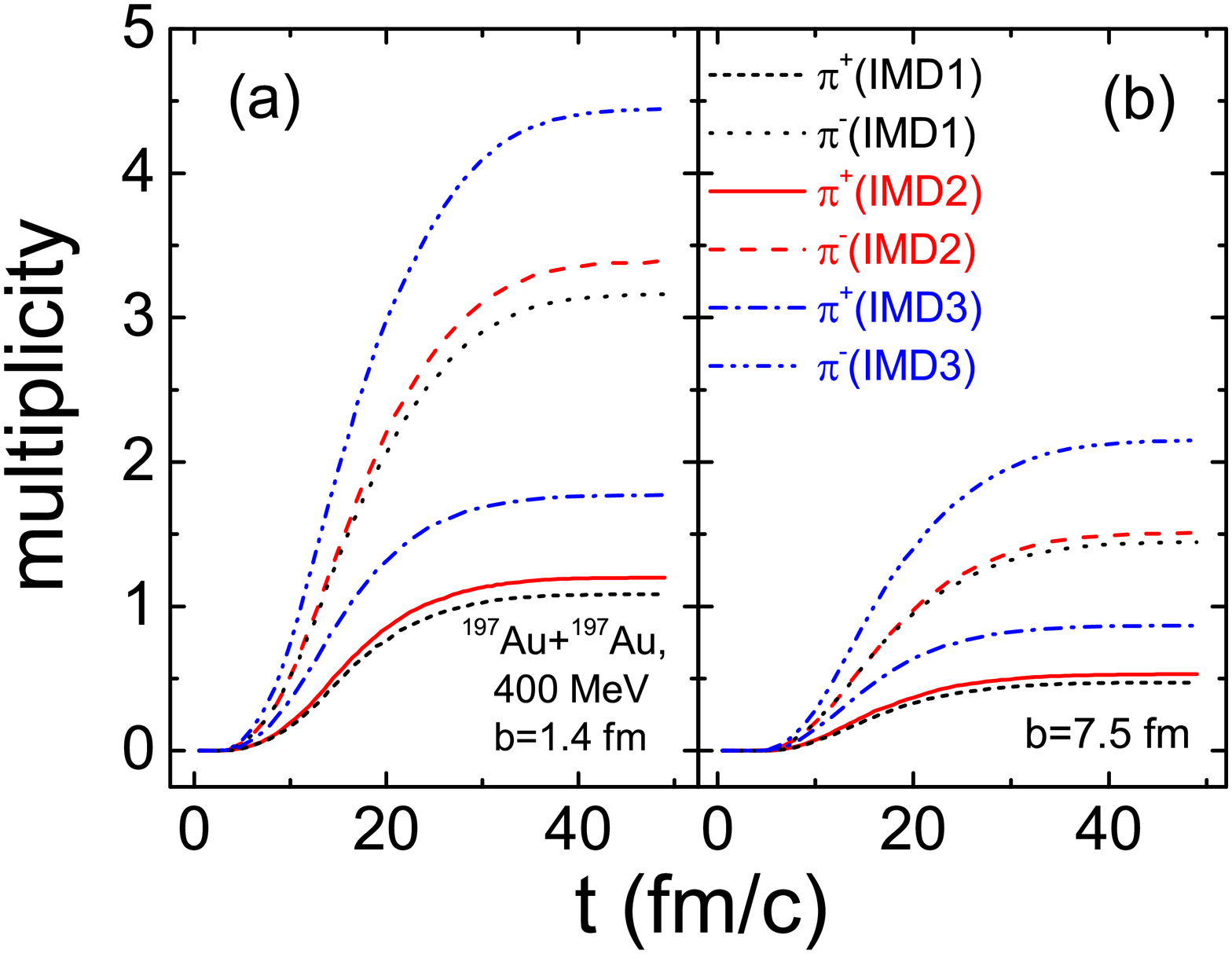}
\caption{(Color online) The yields of $\pi^{-}$, $\pi^{+}$ as a function of time with different initializations of nucleon momentum (IMD1, IMD2, IMD3) in the central (a) and semi-central (b) $^{197}\rm {Au}+^{197}\rm {Au}$ reactions at incident beam energy of 400 MeV/nucleon.} \label{pion}
\end{figure}
\begin{figure}[th]
\centering
\includegraphics[width=0.5\textwidth]{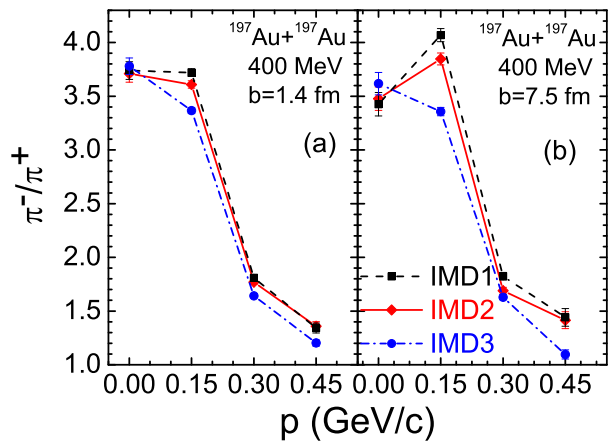}
\caption{(Color online) The ratio of $\pi^{-}$/$\pi^{+}$ as a function of momentum with different initializations of nucleon momentum (IMD1, IMD2, IMD3) in the central (a) and semi-central (b) $^{197}\rm {Au}+^{197}\rm {Au}$ reactions at incident beam energy of 400 MeV/nucleon.} \label{rpion}
\end{figure}
We next move to the discussions of the pion production. Fig.~\ref{pion} shows the yields of $\pi^{-}$, $\pi^{+}$ as a function of time with different initializations of nucleon momentum (IMD1, IMD2, IMD3) in the central (a) and semi-central (b) $^{197}\rm {Au}+^{197}\rm {Au}$ reactions at incident beam energy of 400 MeV/nucleon. It is seen that, whether in the central or in the semi-central collisions, more $\pi$'s are produced with IMD3 than those with IMD1 and IMD2. The numbers of pion production with IMD1 and IMD2 have small difference while the difference becomes larger when comparing them to the case with IMD3. This means that the number of produced $\pi$'s is affected by the fraction of energetic nucleons in the HMT.

To reduce the systematic errors, it is preferable to see the ratio of $\pi^{-}$ over $\pi^{+}$.
Fig.~\ref{rpion} shows the ratio of $\pi^{-}$/$\pi^{+}$ as a function of momentum with different initializations of nucleon momentum (IMD1, IMD2, IMD3) in the central (a) and semi-central (b) $^{197}\rm {Au}+^{197}\rm {Au}$ reactions at incident beam energy of 400 MeV/nucleon. It is seen that the effects of the HMT on the $\pi^{-}$/$\pi^{+}$ ratio are also evident, especially around the Coulomb peak \cite{yong2006}. Around the Coulomb peak, the value of $\pi^{-}$/$\pi^{+}$ ratio with IMD1 is higher than that with IMD2, and the value of $\pi^{-}$/$\pi^{+}$ ratio with IMD2 is higher than that with IMD3. With IMD3, more neutrons and protons are correlated thus proton has larger probability to locate in the HMT in neutron-rich matter. More energetic proton-proton collisions produce more $\pi^{+}$'s, one thus see a low value of the $\pi^{-}$/$\pi^{+}$ ratio. At high momenta, the effects of the HMT on the $\pi^{-}$/$\pi^{+}$ ratio are not very clear, which deserve further study.

\begin{figure}[th]
\centering
\includegraphics[width=0.5\textwidth]{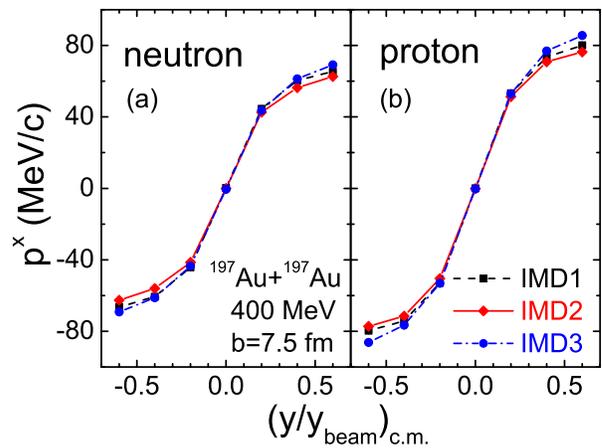}
\caption{(Color online) Nucleon transverse flow $p^{x}$ as a function of rapidity with different initializations of nucleon momentum (IMD1, IMD2, IMD3) in the semi-central $^{197}\rm {Au}+^{197}\rm {Au}$ reaction at incident beam energy of 400 MeV/nucleon.} \label{ntflow}
\end{figure}
\begin{figure}[th]
\centering
\includegraphics[width=0.5\textwidth]{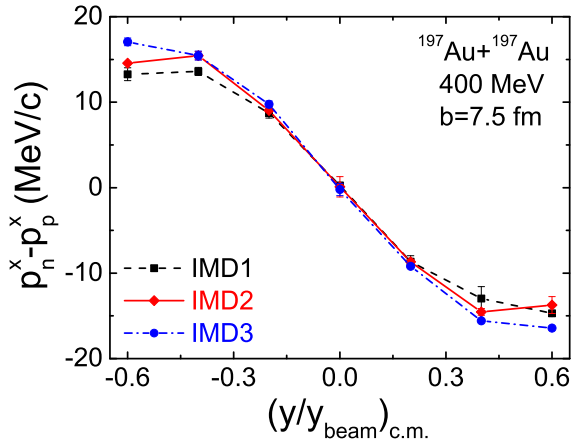}
\caption{(Color online) Relative nucleon transverse flow $p^{x}_{n}$ - $p^{x}_{p}$ as a function of rapidity with different initializations of nucleon momentum (IMD1, IMD2, IMD3) in the semi-central $^{197}\rm {Au}+^{197}\rm {Au}$ reaction at incident beam energy of 400 MeV/nucleon.} \label{nptflow}
\end{figure}
We now turn to the study of the effects of the HMT on the nucleon transverse and elliptic flows. In this study, nucleon transverse flow as a function of rapidity in the semi-central $^{197}$Au+$^{197}$Au reaction at 400 MeV/nucleon reads
$p_{n,p}^{x}=\frac{1}{N_{n,p}(y)}\sum_{i=1}^{N_{n,p}(y)}p_{i(n,p)}^{x}(y)$ with $N_{n,p}(y)$ being the number of neutrons or protons at rapidity $y$ and $p_{i (n,p)}^{x}(y)$ being the transverse momentum (in the direction of impact parameter) of the free nucleon at rapidity $y$.
From Fig.~\ref{ntflow} and Fig.~\ref{nptflow}, both the neutron and proton flows as well as the relative nucleon transverse flow $p^{x}_{n}$ - $p^{x}_{p}$ are all less sensitive to the specific form of the HMT. It seems that the nuclear pressure from dense matter still plays a major role in the transverse flow.

\begin{figure}[th]
\centering
\includegraphics[width=0.5\textwidth]{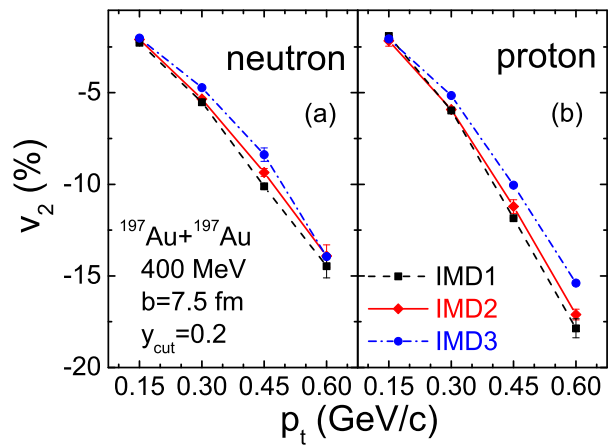}
\caption{(Color online) Nucleon elliptic flow $v_{2}$ as a function of transverse momentum $p_{t}$ ( = $\sqrt{p_{x}^{2}+p_{y}^{2}}$ ) at $(y/y_{beam})_{c.m.}\leq 0.2$ with different initializations of nucleon momentum (IMD1, IMD2, IMD3) in the semi-central $^{197}\rm {Au}+^{197}\rm {Au}$ reaction at incident beam energy of 400 MeV/nucleon.} \label{neflow}
\end{figure}
\begin{figure}[th]
\centering
\includegraphics[width=0.5\textwidth]{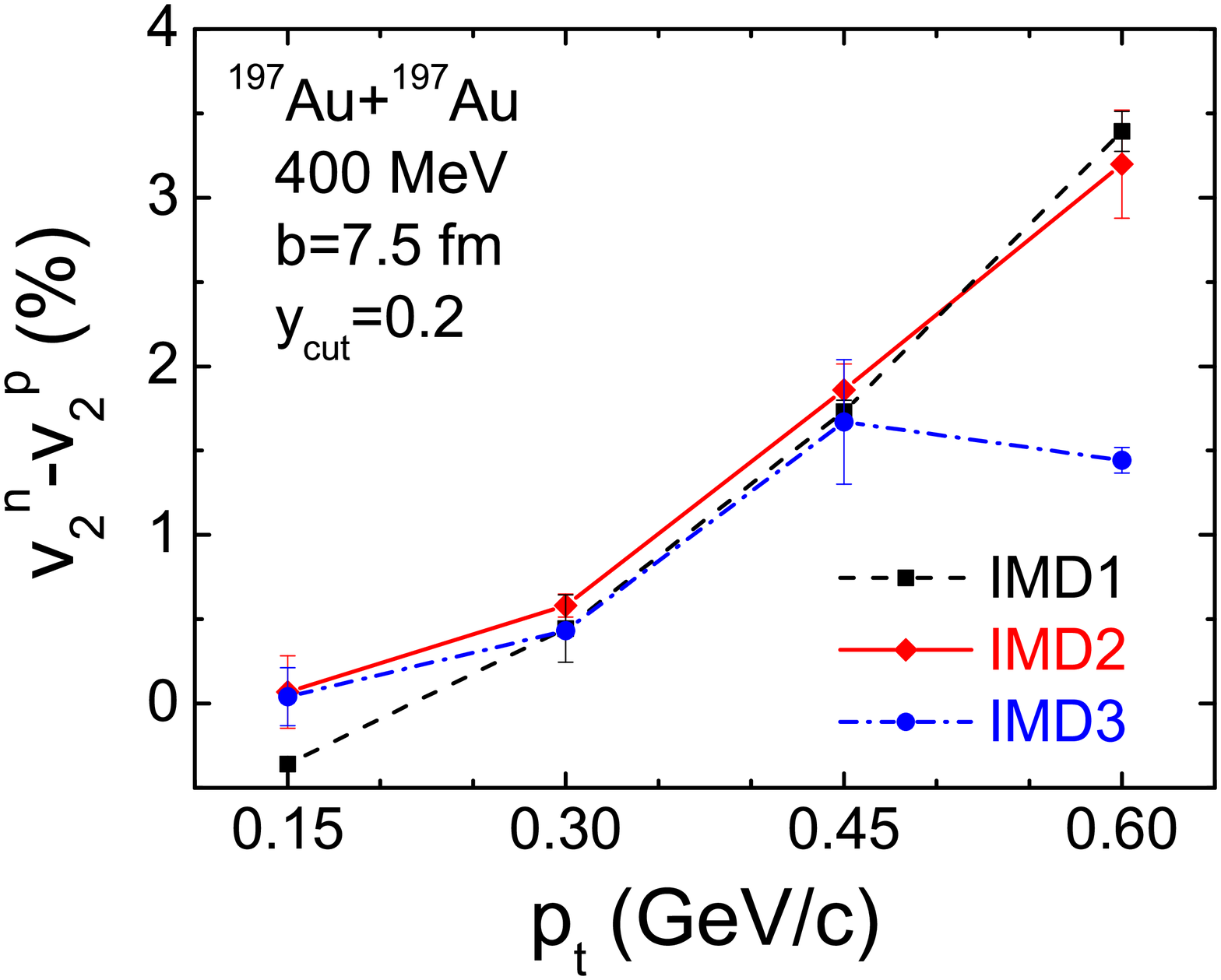}
\caption{(Color online) Relative nucleon elliptic flow $v_{2}^{n}$ - $v_{2}^{p}$ as a function of transverse momentum at $(y/y_{beam})_{c.m.}\leq 0.2$ with different initializations of nucleon momentum (IMD1, IMD2, IMD3) in the semi-central $^{197}\rm {Au}+^{197}\rm {Au}$ reaction at incident beam energy of 400 MeV/nucleon.} \label{npeflow}
\end{figure}
To cancel out nuclear pressure from dense matter to some extend, the nucleon elliptic flow, which reads
$v_{2}=\langle\cos(2\phi)\rangle=\langle\frac{p_{x}^{2}-p_{y}^{2}}{p_{t}^{2}}\rangle
$, seems to be a relative good observable to probe the specific form of the HMT.
Fig.~\ref{neflow} shows the effects of the HMT on the nucleon elliptic flow $v_{2}$ in the semi-central $^{197}\rm {Au}+^{197}\rm {Au}$ reaction at incident beam energy of 400 MeV/nucleon. It is clearly seen that the HMT affects the nucleon elliptic flow, especially for the proton elliptic flow. The larger fraction of nucleons in the HMT (IMD3) corresponds a small anisotropic flow while the small number of HMT's nucleons (IMD1) causes a strong anisotropic flow. Since the proton has larger probability to be in the HMT in neutron-rich matter, the form of the HMT affects the proton anisotropic flow more than the neutron anisotropic flow, especially at high momenta. Fig.~\ref{npeflow} shows the relative nucleon elliptic flow $v_{2}^{n}$ - $v_{2}^{p}$ as a function of transverse momentum $p_{t}$. One can see that the value of the difference of neutron and proton elliptic flows roughly increases as the increase of transverse momentum. It is also seen that at high transverse momenta, the relative nucleon elliptic flow $v_{2}^{n}$ - $v_{2}^{p}$ shows larger effects of the HMT.

\section{summary}

The high-momentum tail of nucleon momentum distribution caused by the nucleon-nucleon short-range correlations in nuclei is confirmed by recent experimental and theoretical studies. Nowadays the specific forms of the high-momentum tail, such as the fraction of high-momentum nucleons, the effective maximum momentum of nucleons in the high-momentum, the neutron to proton ratio of nucleons in the high-momentum tail, etc., deserve further study. Based on the SRC modified transport model, it is shown that the free neutron to proton ratio at high momenta and the yields of $\pi^{-}$ and $\pi^{+}$ in heavy-ion collisions at intermediate energies are preferable to probe the specific form of the high-momentum tail of nucleon momentum distribution in heavy nuclei or nuclear matter. To constrain the symmetry energy by the iso-sensitive observables, one should try to avoid those probes that are also sensitive to the specific form of the HMT.

\section{acknowledgements}

This work is supported in part by the National Natural Science
Foundation of China under Grant Nos. 11775275, 11435014.

\end{document}